
%
\input amstex
\voffset-1cm
\documentstyle{amsppt}
%
%
%
%
%
%
\def\R{\Bbb R }
\def\N{\Bbb N }
\def\P{\Bbb P }

\def\Z{\Bbb Z }
\def\C{\Bbb C }

%
%
\loadbold
\define\np{\vfil\eject}

\redefine\l{\lambda}

\define\a{\alpha}
\define\w{\omega}

\redefine\b{\beta}
\redefine\t{\tau}
\redefine\i{{\,\roman i\,}}
\define\PL{Phys\. Lett\. B}

\define\LMP{Lett\. Math\. Phys\. }
\define\Izv{Math\. USSR Izv\. }

\define\im{\text{Im\kern1.0pt }}
\define\pbar{\overline{\partial}}
\define\rn{\Pi^{(m)}}
\define\Qn#1{\widehat{Q}_{#1}^{(m)}}
\define\Pn#1{\widehat{P}_{#1}^{(m)}}
\define\Tn{T^{(m)}}
\define\Tfn{T_f^{(m)}}
\define\Tgn{T_g^{(m)}}
\define\Tfgn{T_{\{f,g\}}^{(m)}}
\define\La{\frak L_\a}
\define\Lb{\frak L_\b}
\define\Pa{{\Cal P}(M)}

\define\ghn{\Gamma_{hol}(M,L^{m})}

\define\gh{\Gamma_{hol}(M,L)}
\define\Lm{L^{m}}
\define\diff{\text{\it diff}}
\define\Lp{{\roman L}^2(M,L)}
\define\Lq{{\roman L}^2(Q)}
\define\Lqv{{\roman L}^2(Q,\nu)}
\redefine\L{\frak L}

\define\Pii{\Pi^{(1)}}
\define\Qm#1{\widehat{Q}_{#1}^{(m)}}
\define\Qmo{\widehat{Q}^{(m)}}
\define\Pm#1{\widehat{P}_{#1}^{(m)}}
\define\Tm{T^{(m)}}
\define\Tfm{T_f^{(m)}}
\define\Tgm{T_g^{(m)}}

\define\Tma#1{T_{#1}^{(m)}}
\define\ghm{\Gamma_{hol}(M,L^{m})}

\define\Hm{\Cal H^{(m)}}
\define\phm{\varPhi^{(m)}}
\define\phtm{\tilde\varPhi^{(m)}}
\define\Mt{\widetilde{M}}
\define\zb{\overline{z}}
\define\wb{\overline{w}}

\magnification=1200
\NoBlackBoxes
\TagsOnRight
\hfill Mannheimer Manuskripte 147

\hfill University of Freiburg THEP 93/22

(to appear in Commun. Math. Phys.)\hfill hep-th/9309134
\vskip 1cm
\topmatter
\title
Toeplitz Quantization of K\"ahler Manifolds
and $\boldkey g\boldkey l\boldkey(\boldkey N\boldkey)$,
$\boldkey N\boldsymbol \to\boldsymbol\infty$ limits.
\endtitle
\rightheadtext{Toeplitz Quantization of K\"ahler Manifolds
}
\leftheadtext{M. Bordemann, E. Meinrenken  and M. Schlichenmaier}
\author Martin Bordemann, Eckhard Meinrenken\\ and \\
Martin  Schlichenmaier\endauthor
\address
Martin Bordemann,
Department of Physics,
University of Freiburg, Hermann-Herder-Str. 3,
D-79104 Freiburg, Germany
\endaddress
\email
mbor\@ibm.ruf.uni-freiburg.de
\endemail
\address
Eckhard Meinrenken,
Department of Mathematics, M.I.T.,
Cambridge, Mass. 02139, U.S.A.
\endaddress
\email
mein\@math.mit.edu
\endemail
\address
Martin Schlichenmaier,
Department of Mathematics and Computer Science,
University of Mannheim
D-68131 Mannheim, Germany
\endaddress
\email
schlichenmaier\@math.uni-mannheim.de
\endemail
\date September 93
\enddate
\abstract
For general compact  K\"ahler manifolds it is shown that
both Toeplitz quantization and geometric quantization
lead to a well-defined (by operator norm estimates)
classical limit. This generalizes earlier results
of the authors and Klimek and Lesniewski obtained for the
torus and higher genus Riemann surfaces, respectively.
We thereby arrive  at an approximation of the Poisson algebra
by a sequence of finite-dimensional matrix algebras
$gl(N)$, $N\to\infty$.
\endabstract
\endtopmatter
%
%
%
\document
\head
1. Introduction\hfill{ }
\endhead
In a couple of papers titled ``Quantum Riemann Surfaces'' \cite{24} S.~Klimek
and A.~Lesniewski have recently proved a classical limit theorem for the
Poisson algebra of smooth functions on a compact Riemann surface $\Sigma$
of genus $g\ge 2$ (with Petersson K\"ahler structure)
using the Toeplitz quantization procedure:
$$\gather
\lim_{\hbar\to 0}||T^{(1/\hbar)}_f||  =  ||f||_{\infty}, \tag 1-1 \\
\lim_{\hbar\to 0}
||\frac{1}{\hbar}[T^{(1/\hbar)}_f,T^{(1/\hbar)}_g]-\i\,T^{(1/\hbar)}_{\{f,g\}}||
              =  0\ .  \tag 1-2
\endgather $$
Here,
$\frac{1}{\hbar}=1,2,\ldots$
are tensor powers of the quantizing Hermitian line bundle $(L,h)$ over $M$,
and the Toeplitz operators act on the Hilbert space of holomorphic
sections of $L^{1/\hbar}$ as the holomorphic part of
the operator that multiplies sections with $f$.

As usual (1-2) gives the connection between the Poisson bracket of
functions and the commutator of the associated operators and (1-1) prevents
the theory from being empty.
Compared to Berezin's covariant symbols \cite{3} and to the
concept of star products \cite{2,6,9,11}, where the basic idea
is the deformation of the algebraic structure on $C^\infty(M)$ using
$\hbar$ as a formal deformation parameter, the emphasis lies here more
on the approximation of $C^\infty(M)$ by operator algebras in norm sense.
More generally, the estimates (1-1) and (1-2) above can be seen in the
setting of approximating an (infinite-dimensional) Lie algebra
$\L$
by a family
$(\La)$ of metrized Lie algebras indexed by some parameter $\alpha$.

This concept does not only apply to the
classical limit in quantization procedures, but also to other
physical contexts. An important example is the Lie algebra
$\diff_A\Sigma$ of all divergence-free or volume-preserving vector fields
which plays a distinguished r\^ole both in two-dimensional hydrodynamics
\cite{1,13} and in the theory of relativistic membranes
\cite{4,23}. Its relation to  the
Poisson algebra of $\Sigma$ is
that the Poisson algebra is isomorphic (modulo the constant functions) to
the Lie algebra of Hamiltonian vector fields on $\Sigma$, which in turn
is (up to first de Rham cohomology) equal to $\diff_A\Sigma$.
Originally starting
{}from membrane theory (where this limit occurred in a phenomenological way
as approximation of structure constants, see \cite{23}),
an axiomatic treatment of such an approximation scheme
which was called ${\frak L}_{\alpha}$-quasilimit was given in \cite{5}.
Roughly speaking,
quasilimits can be seen as generalized projective limits
with the homomorphisms
$\La\to\Lb$ replaced by certain asymptotic
conditions. Apart from several examples the paper \cite{5} also contains
the relation to classical limits via geometric quantization on
compact K\"ahler
manifolds and the proof
of (1-1) and (1-2) for the Poisson algebra
on the $2n$-torus using theta functions (with characteristics).
The above Toeplitz operators $T^{(1/\hbar)}_f$ were replaced by the
operators of geometric quantization $Q^{(1/\hbar)}_f$, but the asymptotic
results are equivalent according to Tuynman's relation
$Q^{(1/\hbar)}_f=\i T^{(1/\hbar)}_{f-(\hbar/2)\Delta f}$.

The aim of this paper is to generalize the classical limit for Toeplitz
quantization of the above Riemann surfaces to the general compact
K\"ahler case (the ``quantum K\"ahler manifolds''),
i.e\.
to prove (1-1) and (1-2) in this context
and to use them to show the following theorem  (conjectured in \cite{5}):
\proclaim{Theorem }
Let $(M,\w)$ be a quantizable compact K\"ahler manifold, $\w$ the
K\"ahler form,
$\Pa$ the Poisson algebra of real valued $C^\infty-$functions
with respect to $\w$,
$L$ the quantum line bundle, and $L^m$ its $m^{th}$ tensor power.
Let $\w$ be rescaled (by multiplying it with a positive integer)
 in such a way that
$L$ is very ample.
Then, with respect to
the maps
$\ f\to \i m\,T_f^{(m)}\ $and
$\ f\to m\,Q_f^{(m)}\ $
the Poisson algebra $\Pa$ is
a $\ u(\dim \ghn)-$quasilimit ($m\to\infty$) in both cases.
\endproclaim
The technical details entering the hypotheses of this theorem will be
explained below. We believe that one can probably
dispense with the condition that the bundle is very ample
(i.e\. avoid the rescaling).

The proof is
largely based on the theory of generalized Toeplitz structures developed
in the mid-seventies by L.~Boutet de Monvel,
V.~Guillemin, and J.~Sj\"ostrand in the framework of microlocal analysis
\cite{7,8,18}.
In fact, the estimate (1-2)
is an easy consequence of the symbol calculus for generalized Toeplitz
operators, whereas the innocent looking (1-1) requires more
efforts.

Let us give a rough outline of the arguments.
Denote by $U$ the dual line bundle to $L$, along with its Hermitian
fibre metric, and by $Q$ the unit disc bundle. Sections of $L^m$
can be identified with functions on $Q$ satisfying appropriate
equivariance
conditions. In this way, the direct sum of the spaces of holomorphic
sections of $L^m$ gets identified with a Hilbert subspace of
$\Lq$, called generalized Hardy space.
As shown in \cite{7,8,18}, the orthogonal projector onto the Hardy
space has good microlocal properties, and renders a ring of generalized
Toeplitz operators on $\Lq$ having properties similar to
pseudo-differential operators.
On the other hand, the spaces of holomorphic sections of $L^m$ can
be recovered using Fourier decomposition with respect to the natural
circle action on the Hardy space, and the symbol calculus for the
generalized Toeplitz operators gives the desired approximation results
for the original problem.

The paper is organized as follows.
In Section 2 we  recall the notion of $\La-$quasi\-li\-mit
and describe its relation to geometric quantization for the
convenience  of the reader and to fix notation.
In Section 3 we discuss the above theorem
for projective K\"ahler manifolds and Riemann surfaces.
In Section 4 we formulate
the basic asymptotic results for partial Toeplitz operators
(the Equations (1-1) and (1-2) above)  and explain why this implies
the main theorem.
Their proof is given in section 5.

\head
2. $\La-$quasi-limits and geometric quantization
\hfill { }\
\endhead
We recall from \cite{5} the definition of an $\La-$quasilimit.
Let $\ (\L,[\;\;,\;])\ $ be a real or a complex Lie algebra
and $\ (\La,[\;\;,\;]_\a))_{\a\in I}\ $ a family of real resp\. complex Lie
algebras
with index set $I$ either $\N$ or other suitable subsets of $\R$\ .
Let the  Lie algebras $\La$ be equipped with metrics $\ d_\a\ $
(in our cases they are all coming from a norm) and let
$\ (p_\a:\ \L\to\ \La)_{\a\in I}\ $
be a family of linear maps.
\definition {Definition 2.1}
$\ (\La,[\;\;,\;]_\a))_{\a\in I}\ $ is called an {\it approximating
sequence} for $\ (\L,[\;\;,\;])\ $ and $\ (\L,[\;\;,\;])\ $ is called an
$\La-$quasilimit induced by $\ (p_\a:\ \L\to\ \La)_{\a\in I}\ $
if
\roster
\item
all $p_\a$ for $\a\gg 0$ are surjective,
\item
if for all $x,y\in \L$ we have
$\ d_\a(p_\a(x),p_\a(y))\ \to\  0$, for $\ \a\to\infty$
 then $\ x=y\ $,
\item
for all $x,y\in \L$ we have
$\ d_\a(p_\a([x,y]),[p_\a(x),p_\a(y)]_\a)\ \to\  0,\ $
 for $\ \a\to\infty$.
\endroster
\enddefinition
{}From  (2) it follows that an element which is asymptotically
zero is already zero and from (3) it follows that there is only one
Lie product on $L$ which is compatible with a given
approximating sequence and a given system of maps $(p_\a)$.
For examples
we refer to \cite{5\rm, Sect.~3}.

As was pointed out to us by J.~B.~Bost this definition is related
to the notion of continuous fields of $C^*$-algebras
as introduced in \cite{12}.
\vskip 0.4cm
Let $M$ be a compact K\"ahler manifold of complex dimension $n$
with K\"ahler form $\w$.
In particular, $\ (M,
\w)\ $ is a symplectic manifold.
For every smooth function $f$ on $M$ the Hamiltonian vector field
$X_f$ is defined by $\ i_{X_f}(\w)=df\ $.
Let  $\ \Pa\ $ be the Lie algebra of smooth functions on $M$ with
the Lie bracket
$$ \{f,g\}:=df(X_g)=\w(X_f,X_g)\ .\tag 2-1$$
Now let $(M,\w)$ be a quantizable manifold and $L$ be a holomorphic
quantum line bundle with fiber metric $h$ and compatible covariant
derivative $\nabla$.
For the explanation of the above terms we refer to
\cite{5\rm, Sect.~4} for a quick review, resp\. to \cite{34},
\cite{31}, \cite{32}
for detailed information.

The condition for $L$ to be a quantum line bundle for $(M,\w)$ says
that the curvature of $L$ is essentially equal to the symplectic
form. More precisely for every pair of vector fields $\ X,Y\ $
we have the prequantum condition
$$F(X,Y)=\nabla_X\nabla_Y-\nabla_Y\nabla_X-
\nabla_{[X,Y]}=
-\i \w(X,Y)\ .\tag 2-2$$
By this definition $L$ is a positive line bundle.
According to  Kodaira's embedding theorem some tensor power
$\Lm$ is ``very ample'', i.e\. one gets a holomorphic embedding
of $M$ into a projective space using the holomorphic sections of
$\Lm$. After the
choice of  a basis $\varphi_0,\dots, \varphi_N$ of $\ghn$
this embedding is given as
$$ M\to\P^N,\qquad x\mapsto (\varphi_0(x):\varphi_1(x):\ldots:\varphi_N(x))\ .
$$
Chow's theorem says that
$M$ is in fact a projective algebraic manifold
\cite{29,p.60}.

For every smooth  function
$f$ on $M$ the following prequantum operator $P_f$ acting on the
complex vector space $\Gamma(M,L)$ of all smooth global sections of
$L$ is formed
$\  P_f:=-\nabla_{X_f}+\i f\cdot 1\ $.
This defines a map
$$P:\Cal P(M)\to gl(\Gamma(M,L)),\quad
f\mapsto P_f\ .$$
By the prequantum condition (2-2)
the map $P$ is  an injective Lie algebra homomorphism.
Let $\Omega=\frac{1}{n!}\omega^n$ denote the symplectic volume form on $M$,
and define
the prequantum Hilbert space $\Lp$
as the completion of $\Gamma(M,L)$ with respect to
the scalar product
$$< \varphi\,|\,\psi>:=\int_Mh(\varphi,\psi)\Omega\ . \tag 2-3$$

With respect to this scalar product
$P_f$ becomes an antihermitian operator of
$\Gamma(M,L)$ for real valued $f$.

A second step in the geometric quantization scheme is the
choice of a polarization. The canonical concept for K\"ahler manifolds
is the separation into holomorphic and
anti-holomorphic directions, called K\"ahler polarization. The
quantum Hilbert space is the subspace
$\ \gh\ $ of holomorphic sections in $\Lp$.
Due to compactness of $M$ the space $\ \gh\ $
is always finite dimensional.
The quantum operator $Q_f$ is
defined as $\ Q_f:=\Pii\circ P_f\circ \Pii\ $,
where $\ \Pii:\Lp\to \gh\ $  denotes orthogonal  projection.
The map $\ Q:f\mapsto Q_f\ $ is a linear map from $\Pa$ to the
finite dimensional Lie algebra $\ u(\gh)\ $ of antihermitian
operators in $\gh$.

In this paper, however, we will be more concerned with  Toeplitz
quantization, defined as follows.
For  $f\in\Pa$ the corresponding   Toeplitz operator on $\gh$ is the
operator of multiplication $M_f$ by $f$ followed by orthogonal  projection
back to $\gh$,
$$T_f:=T(f):=\Pii\circ M_f\circ \Pii\ .\tag 2-4$$
According to a result of Tuynman \cite{32} (see also \cite{5\rm, Prop.~4.1})
one has
$$Q_f=\i\; T(f-\frac 12\Delta f)\ .\tag 2-5$$
Here $\Delta$ is the Laplacian on functions calculated with
respect to the Riemannian metric $g$ coming from $\w$.
\medskip
To obtain a family of finite dimensional Lie algebras
associated to $\Pa$
 we consider
everything for the $m^{th}$ tensor power $L^m:=L^{\otimes m}$
of the quantum line bundle $L$ for $m\in\N$.
The quantum Hilbert space is thus $\ghn$, with scalar product
$$ < \varphi|\; \psi>:=\int_M h^m(\varphi,\psi)\,\Omega,\quad
h^m:=h\otimes\cdots \otimes h\ \quad (m \text{ factors})\ ,\tag 2-6$$
and
the prequantum operators $\ P_f^{(m)}\ $
define a representation of $(\Pa,m\cdot\w)$.
In order to render  a representation of $(\Pa,\w)$
they have to be rescaled
to $\ \Pn f:=m P_f^{(m)}
=-\nabla_{X_f}^{(m)}+\i mf$.
The rescaled quantum operators
are given as
$$\Qn f:=\rn\circ\Pn f\circ\rn,\tag 2-7$$
with $\rn$ the corresponding projection map.
By  Equation~(2-5)  one has
$$\Qn f=\i m\;\Tn(f-\frac 1{2m}\Delta f)\ .\tag 2-8$$
Note that neither  $\Tn$
nor the Laplacian are rescaled.

For the elements in $gl(\ghn)$ we
take the rescaled
norm
$$||A||_m:=\frac 1m\sup_{\varphi\ne 0}
\frac {||A\,\varphi||}{||\varphi||},
\tag 2-9$$
and $\ ||..||\ $ the operator norm.

For $\varphi,\psi\in\ghn$ we obtain
$$< \varphi|\;T_f^{(m)} \psi>=
< \rn \varphi|\;f\cdot \rn \psi>=
< \varphi|\;f\cdot  \psi>=\int_M fh^m(\varphi,\psi)\,\Omega\ .\tag 2-10$$
The settings for $m\in\N$ with $m\to\infty$,

$$
\gather
(\Pa,\{\;{\;},\;\})\quad\to\quad
(\ u(\ghn)\ ,\ [\;{\;},\;]\ ,\ ||..||_m\ ),\
p_m:f\mapsto \Qn f,\tag 2-11\\
(\Pa,\{\;{\;},\;\})\quad\to\quad
(\ u(\ghn)\ ,\ [\;{\;},\;]\ ,\ ||..||_m\ ),\
p_m:f\mapsto \i m\cdot\Tm_f,\tag 2-12
\endgather  $$
are exactly  the settings examined in the scheme of $\La-$quasilimits.
That $m^{-1}$ is likely to play the role of $\hbar$ is already indicated
by the formula for the
dimension of $\ghn$. Indeed, the Hirzebruch--Riemann--Roch  theorem
says that for $m$ large, this dimension is a polynomial in $m$
with leading term
$$\dim\ghn=\frac{m^n}{(2\pi)^n}
\text{vol} (M) +O(m^{n-1}),\tag 2-13$$
where $\text{vol} (M)$ is the symplectic volume. But this is just what
is to be expected from the uncertainty relation.

\np
\head
3. The approximation theorem \hfill { }
\endhead
The following theorem will be proved in the remaining Sections 4 and 5.
\proclaim{Theorem 3.1}
Let $(M,\w)$ be a quantizable compact K\"ahler manifold,
$\ \Pa\ $ the Poisson algebra of real valued $C^\infty-$functions
with respect to $\w$,
$L$ the quantum line bundle, and $L^m$ its $m^{th}$ tensor power.
Let $\w$ be rescaled (by multiplying it with a positive integer)
 in such a way that
$L$ is very ample.
Then, with respect to both settings (2-11) and (2-12)
$\Pa$ is
a $\ u(\dim \ghn)-$quasilimit ($m\to\infty$).
\endproclaim

Let us illustrate the theorem by two important
special classes of examples:
The first class consists of the projective K\"ahler submanifolds.
For the $N$-dimensional projective space $\P^N$ the Fubini-Study
fundamental form $\w_{FS}$ is defined as $$\w_{FS}:=
\i\frac
{(1+|w|^2)\sum_{i=1}^Ndw_i\wedge
d\wb_i-\sum_{i,j=1}^N\wb_iw_jdw_i\wedge d\wb_j} {{(1+|w|^2)}^2}
\tag 3-1$$
with respect to the local coordinates $w_i=z_i/z_0$, $i=1,\ldots,N$ on
the coordinate chart where the homogeneous coordinate $z_0\ne 0$ (see
for example \cite{33}).
It defines the standard K\"ahler form on $\P^N$
and it is up to the scalar factor $-\i$ the curvature form of the
hyperplane bundle $H$.  Hence, $H$ is an associated quantum line
bundle.

Now let $\ i:M\hookrightarrow \P^N\ $  be a projective K\"ahler
submanifold of dimension $n$.  The pullback $L=i^*(H)$ (resp\. the
restriction) of the hyperplane bundle $H$ is a quantum line bundle
associated to the pullback $i^*(\w_{FS})$ which is the
K\"ahler form of $M$.  The space of global
holomorphic sections of $L^m$ is generated by the restrictions of the
homogeneous polynomials of degree $m$ in $N+1$ variables.  Note that
 they
are in general not linearly independent when restricted to $M$.
Formula (2-13) is the Hilbert polynomial of $M$, i.e.
$n! \frac{\text{vol} (M)}{(2\pi)^n}$ (which is a positive integer)
is equal the degree of $M$ considered as a projective submanifold.

The second class of examples are Riemann surfaces with their
``standard'' K\"ahler forms.  For the rest of this section let $M$ be a
compact Riemann surface with fixed complex structure.
Depending on the type of the simply connected universal
covering $\widetilde{M}$ of $M$
the
classes of Riemann surfaces can be divided into three
subclasses (see \cite{15}, \cite{29}).
\subhead
Case 1
\endsubhead
Here $\Mt=\P^1$, the projective line over $\C$, resp\.
the sphere $S^2$. In this case $M\cong \Mt=\P^1$.
This isomorphism
like all
other isomorphisms appearing in the following
is an analytic isomorphism.
We use the standard covering of
$\P^1$ by the open sets
$U_0$ and $U_1$, $U_0\cong U_1\cong\C$
$$U_0:=\{(z_0:z_1)\mid z_0\ne 0\},\quad
U_1:=\{(z_0:z_1)\mid z_1\ne 0\}\ .$$
We take $\ z=z_1/z_0\ $ as coordinate for $U_0$, and
$\ w=z_0/z_1\ $ as coordinate for $U_1$. The transition function
is given as $\ w(z)=1/z\ $.
In the following we will describe every object by local functions in
$U_0$.
The K\"ahler form (3-1) specializes to
$$\omega_0(z)=\frac {\i}{(1+z\zb)^2}\; dz\wedge d\zb\ .\tag 3-2$$
The corresponding quantum line bundle is the hyperplane bundle
$L_0$ with transition function $1/z$.
Its global holomorphic sections are the elements of the
vector space
$\ \langle 1,z\rangle_{\C}\ $.
For the tensor powers $\ L_0^{m}:=L_0^{\otimes m}\ $ we obtain
(for example by using the theorem of Riemann Roch \cite{29})
$\ \dim \Gamma_{hol}(\P^1,L_0^m)=m+1\ $.
A basis is given by $\ 1,z^1,z^2,\ldots,z^m\ $.
\subhead
Case 2
\endsubhead
 $\widetilde{M}=\C$. In this case $M$ is a one dimensional complex
torus, e.g\. $M\cong \C/\Gamma\ $ where $\ \Gamma
=\langle 1,\t\rangle_{\Z}\  $
($\im \t>0 $) is a two
dimensional lattice in $\C$. The genus of $M$ is equal to 1 and the
K\"ahler form is given by
$$\omega_1(z)=\frac {\i\pi}{\im \t} dz\wedge d\zb\ .\tag 3-3$$
Here $z$ is the coordinate on the covering. A corresponding
quantum line bundle is  the theta line bundle $L_1$ of degree 1.
It depends on the  complex structure of $M$, e.g. on $\t$.
Its space of global sections is one dimensional and a basis element is
given by the Riemann theta function (see \cite{5\rm, Sect.~5}).
By the Riemann Roch theorem we get
for the tensor powers $L_1^m$
$\ \dim\Gamma_{hol}(M,L_1^m)=m\ $.
These spaces are generated by the theta functions with characteristics.
Of course, $L_1$ is only ample. But $L^{\otimes 3}$ will be very ample
\cite{17}.
\subhead
Case 3
\endsubhead
$\widetilde{M}= E$ with
$\ E:=\{z\in\C\mid |z|<1\}\ $ the open unit disc.
There exists a Fuchsian group $D$, i.e\. a discrete subgroup
satisfying some additional conditions (see \cite{15})
of
$$SU(1,1)\ :=\ \{\pmatrix a&b\\ \overline{b} &\overline{a}\endpmatrix\in
GL(2,\C)\mid |a|^2-|b|^2=1\} ,$$
such that
$\ M\cong E/D\ $ (analytically).
Here the elements $R\in SU(1,1)$ operate by
fractional linear transformations
$$z\mapsto R(z):=\frac {a z + b} {\overline{a} z + \overline{b}}$$
on $E$. This situation could equivalently be described by the
upper half plane and the group $\ SL(2,\R)\ $.
As K\"ahler form on $E$ we take
$$\w=\frac {2\i}{(1-z\zb)^2}dz\wedge d\zb\ .\tag 3-4$$
Because $\ R'(z)=(\overline{b} z+\overline{a})^{-2}\ $ we obtain
$\w(R(z))=\w(z)\ $. Hence (3-4) is invariant
under $SU(1,1)$  and defines a K\"ahler form
$\ \w_g\ $ on $M$.

An associated quantum line bundle $L_g$ is the canonical line bundle
$K$ (i.e\. the line bundle whose local sections are the
local holomorphic differentials). Again, $K$ resp\. $L_g$
depends on the complex structure, i.e\. on the group $D$.
For generic Riemann surfaces of genus $g > 2$ the bundle $L_g$ is
already very ample. In any case $L^{\otimes 3}_g$ will be very ample
\cite{27}.

The bundles $L_g^m$ are the $m-$canonical bundles. By the
theorem of Riemann Roch we obtain
$$\dim \Gamma_{hol}(M,L_g^m)\ =\ \cases g,&m=1,\\
                  (2m-1)(g-1),&m\ge 2\ .
\endcases
$$
As in the $g=1$ case the sections can be identified with functions
on the covering space $E$ which behave suitably under the operation
of the group $D$.
A holomorphic function $f$ on $E$ is called an automorphic form
of weight
\footnote{
The definition of weight varies in literature.
Our weight $2k$ is sometimes called weight $k$
or dimension $-2k$,...
} $\ 2k\ $
for the group $D$ if for every
$\
R=\pmatrix a&b\\ \overline{b} &\overline{a}\endpmatrix\in D$,
$$f(R(z))=(\overline{b} z+\overline{a})^{2k}\cdot f(z)=
(R'(z))^{-k}\cdot f(z)\ .$$
{}From the definition it is clear
that $\ f(z)(dz)^k=f(R(z))(d(R(z)))^k\ $.
Hence, such an automorphic form of weight $2k$ defines a section
of $L_g^k$. Conversely, every such section defines by pullback
an automorphic form on $E$.

Note that in all the above cases the theorem also holds
without the ``very ample'' condition,
see \cite{5},\cite{24}.

\head
4. Approximation and Toeplitz operators
\endhead
Let $(M,\omega)$ be a quantizable compact  K\"ahler manifold and $L$ some
quantum line bundle with metric $h$ over $M$.
We assume $L$ to be  very ample.
Let $\Hm
=\ghm$ be the Hilbert space
of holomorphic sections in $L^m$, with scalar product
(2-6).
Recall the relation (2-8)
between the quantum operators
and the multiplication (Toeplitz) operators.
We will show that  Theorem 3.1 will follow from
Theorem~4.1 and Theorem~4.2 below. In Section 5 we will prove these theorems.
\vskip 0.4cm
First we will show that the
surjectivity (property (1) in Definition~2.1) is always true, due to
the following propositions.

\proclaim{Proposition 4.1}
The canonical linear mapping
$$s^{(m)}:End(\Hm)\to C^\infty(M)
\quad\text{defined by }\quad
s^{(m)}(|\psi><\varphi|):=h^{(m)}(\psi,\varphi )\ ,\tag 4-1$$
is an injection.
\endproclaim
\demo{Proof}
Let $e_1,e_2,\ldots,e_d$ be a basis for $\Hm$. In a local complex chart
$(V,z)$ these sections are represented by holomorphic functions
$e_i(z)$. In this chart the
$d^2$ sections
$\ s^{(m)}(|e_i><e_j|) \ $ are given by the $d^2$ functions
$\ h(z)\overline{e_i(z)}e_j(z)$,
where $h$ is some fixed positive function.
Suppose that
$$\sum a_{ij}h(z)\overline{e_i(z)}e_j(z)=0\ ,$$
for some $a_{ij}\in\C$.
After dividing by $h$, this can be analytically extended to $V\times V$:
$$\sum a_{ij}\overline{e_i(z)}e_j(w)=0\quad\forall z,w\in V\ .$$
It follows that $a_{ij}=0.$
\qed
\enddemo
\proclaim{Proposition 4.2}
The linear mappings $\Tm$ and $\Qmo$:
 $C^\infty(M)\to End(\Hm)$ are surjections.
\endproclaim
\demo{Proof}
For all $f\in C^\infty(M)$ and $A\in End(\Hm)$, one has
for the Hilbert-Schmidt scalar product
$$\langle A|\;\Tfm\rangle=
tr(A^*\cdot \Tfm)=\int_Mf(x)s^{(m)}(A^*)(x)\Omega(x)
=\langle s^{(m)}(A),f \rangle_{L^2}
\ .\tag 4-2$$
Suppose that $A$ is orthogonal to the range of  $\Tm$.
Then both sides of (4-2) vanish for all $f$, i.e\.
$s^{(m)}(A)=0$.
According to Prop.~4.1, this implies $A=0$, hence  $\Tm$
is surjective.
The analogous result for $\Qmo$ follows from
$\ \Qmo=m\Tm\circ(1-\frac 1{2m}\Delta), $
since $(1-\frac 1{2m}\Delta)$ is positive and elliptic.
Hence, for every  $g\in C^\infty $ there is a  $f\in C^\infty$
with $(1-\frac 1{2m}\Delta)f=g$.\qed
\enddemo
\vskip 0.5cm
\proclaim{Theorem 4.1}
For every  $\ f\in \Pa\ $  there is  some $C>0$ such that
$$||f||_\infty-\frac Cm\le||\Tfm||\le
||f||_\infty\quad
\text{as}\quad m\to\infty\ .\tag 4-3$$
Here $||f||_\infty$ is the sup-norm of $f$ on $M$ and
$||\Tfn||$ is the operator norm on $\Hm$.
In particular,
$$\lim_{m\to\infty}||\Tfn||=||f||_{\infty}\ .\tag 4-4$$
\endproclaim
\proclaim{Theorem 4.2}
For all  $f,g\in \Pa\ $,
$$
|| m[\Tfn,\Tgn]-\i \Tfgn||\quad=O(m^{-1})\quad
\text{as }m\to\infty
\ .\tag 4-5$$
\endproclaim
\noindent
{}From both theorems it follows immediately
$$\lim_{m\to\infty} ||\;[\Tfn,\Tgn]\;||=0\ .\tag 4-6$$
\medskip
\demo{Proof of Theorem 3.1}
The required surjectivity is just Prop.~4.2.
Obviously for the  assignment
$f\to \i m\cdot\Tfm$ , by (4-4) and (4-5) the
 remaining two conditions are fulfilled.
Hence for the setting (2-12) Theorem~3.1 is true.
(Note, we use the rescaled operator norm $||..||_m$.)
Using the relations (2-8) which connects the quantum operator
with the Toeplitz operator it is easy to check
(using (4-6) ) that
$$
\gather \lim_{m\to\infty}||\Qn f||_m=||f||_\infty,
\tag 4-7\\
 \lim_{m\to\infty}||\;[\Qn f,\Qn g]- \Qn {\{f,g\}}||_m= 0\ .
\tag 4-8
\endgather$$
Hence, we obtain Theorem~3.1 also for the setting (2-11).
\qed\enddemo
\remark{Remark}
In the case of  Riemann surfaces
Theorem 4.1 and 4.2 have been already proved
by tedious calculations.
Klimek and Lesniewski \cite{24}  did the case of genus $g\ge 2$.
Our Theorem 4.1 corresponds to \cite{24,\rm II.}, Theorem A and
Theorem 4.2 corresponds to  \cite{24,\rm II.} Corollary to Theorem B.
Note that we defined our Poisson bracket (2-1) with the opposite
sign of the bracket used in \cite{24}.
The case $g=1$ has been done by the authors in \cite{5} as a special
case of $n-$dimensional complex algebraic tori.
The authors (unpublished) also did the case $g=0$ using asymptotics of
binomials (Stirling formula, etc\.).
\endremark
\vskip 0.5cm

Before we prove these theorems in Section 5 for the general setting
we will give a more elementary proof of Theorem~4.1
for the first class of
examples, the projective K\"ahler manifolds
$M$.
Let $i: M \hookrightarrow\P^N$ be a nonsingular projective variety, and
$\pi: U\to M$ be the restriction of the tautological line
bundle of $\P^N$ to $M$ with its induced Hermitian structure $k$.
The bundle $U$ is the dual of $L$, the pullback of the hyperplane bundle $H$,
i.e\.
$U=L^*=i^*(H^*)$.
Then $L$ is a quantum bundle
of $(M, \omega)$, where $\omega$ is the pullback of the
Fubini-Study form of $\P^N$. Using the scalar product
on $\C^{(N+1)}$ the metric $k$ extends to
 a function on $U\times U$, holomorphic
in the second argument and anti-holomorphic in the first.
In particular, the Calabi (diastatic) function
\cite{9},\cite{10}
$$D:M\times M\to
\R_{\ge 0}\cup\{\infty\},\quad
D(\pi(\lambda),\pi(\mu))=-\log
|k(\lambda,\mu)|^2\tag 4-9$$
(where we have to choose $\lambda$ and $\mu$ with
$k(\lambda,\lambda)=k(\mu,\mu)=1$
representing  the points of $M$) is well-defined and
vanishes only along the diagonal.
\demo{Proof of Thm~4.1 for these cases}
The second inequality follows directly from the definition (2-4) of
$T_f$.
To proof the first, let $x_0\in M$ be a point where $|f|$ assumes
its supremum, and fix a $\lambda_0\in \pi^{-1}(x_0)$ with
$k(\lambda_0,\lambda_0)=1$.
Identifying holomorphic sections $\phm$ of
$U^{-m}=L^m$ with  holomorphic
functions $\phtm:U\to \C$
which are equivariant
(i.e\. which obey $\phtm(\a v)=\a^m\phtm(v)$),
we define a sequence
$\phm\in\Hm$ by setting
$$\phtm(\lambda)=k(\lambda_0,\lambda)^m
\ .$$
Note that $h^m(\phm,\phm)(x)=\exp(-mD(x_0,x))$.
(Recall, we chose $\lambda$ such that $k(\lambda,\lambda)=1$.)
Hence, using
 Cauchy-Schwartz's inequality
$$\gather
||\Tfm||\ge \frac {||\Tfm\phm||}{||\phm||}
\ge \frac {|<\phm|\Tfm|\phm>|}{<\phm|\phm>}
\\=
\frac {|\int_Mf(x)h^m(\phm,\phm)(x)\Omega(x)|}
{\int_Mh^m(\phm, \phm)(x)\Omega(x)}=
\frac {|\int_Mf(x)e^{-mD(x_0,x)}\Omega(x)|}
{\int_Me^{-mD(x_0,x)}\Omega(x)}\ .
\endgather
$$
Both integrands vanish exponentially
(with respect to $m\to\infty$) outside $x=x_0$.
Moreover, as a function of $x$ the Calabi function
has a nondegenerate critical point at $x=x_0$, i.e\. one
can apply the stationary phase theorem \cite{22}
to both integrals to conclude that
$$||\Tfm||\ge |f(x_0)|+O(m^{-1})=
||f||_\infty+O(m^{-1}) \ .\qed$$
\enddemo

\head
5. Proofs of Theorem 4.1 and 4.2
\endhead
\noindent

The proofs will follow from the theory of ``global''
Toeplitz operators as developed by
L. Boutet de Monvel and V. Guillemin \cite{7}.
Let us review the necessary pre\-re\-qui\-si\-tes from their book.
Let $(M,\w)$ be an $n-$dimensional
 K\"ahler manifold, $\ (U,k):=(L^*,h^{-1})\ $
be the dual of the quantum line bundle as above, and
$$\hat k:U\to \R_{\ge 0},\quad \hat k(\l)=k(\l,\l)\ .$$
Let $\ Q=\hat k^{-1}(1)\ $ be the unit circle bundle.

It is known (see e.g\. \cite{6}) that the 2-form
$\ \i\partial\overline{\partial}\;\hat k\ $ on $U$ is K\"ahler off the
zero section. In particular, the unit disc bundle is strictly
pseudoconvex.

The natural circle action makes $Q$  into a principal $S^1$ bundle
$\ \tau:Q\to M\ $, and the tensor powers of $U$ may be viewed as
associated bundles.
 Let $\i\a\in \i\Omega^1(Q)$ be the
$\frak u(1)-$valued connection 1-form
corresponding to the Hermitian linear connection $\nabla$ on $U$.
($\a$ is the restriction of the  1-form
$\ \frac 1{2\i}(\partial\hat k-\pbar\hat k) \ $ to the
circle bundle.)
According to the prequantum condition,
$\ d\alpha=\tau^*\w$, and $\nu=\frac 1{2\pi}\tau^*\Omega\wedge \a\ $ is
a volume form on $Q$. The generalized Hardy space $\Cal H$ is defined
as the closure in $\Lqv$ of the subspace of all $f\in C^\infty(Q)$
that extend to holomorphic functions on the disc bundle.
$\Cal H$ is preserved under the
circle action and thus splits into a (completed) direct sum
$\ \Cal H=\sum_{m=0}^\infty\Hm\ $, where
$c\in S^1$ acts on $\Hm$ by multiplication with
$c^m$.
Under the identification of sections of $L^m$ with functions on $Q$
satisfying the the equivariance condition
$\ \phi(c\l)=c^m\phi(\l),(c\in S^1)\ $, the Fourier sectors
$\Hm$ coincide with the Hilbert spaces defined in Section 4.
The orthogonal projector $\Pi:\Lq\to\Cal H$
is called the generalized Szeg\"o projector.

We shall asume that $L$ is very ample, i.e\. that
$M$ can be embedded into some projective space $\P^N$ via the
global holomorphic sections of $L$.
In particular, $L$ is the restriction (pullback) of the hyperplane
bundle and $U$ is the restriction of the tautological bundle.
Away from the zero section the latter and hence $U$ can be embedded
into $\C^{N+1}$.
The image of $U$ is an affine cone, hence a Stein variety (with singularity
at 0 coming from the collapse of the
zero section).
Under this condition $\Pi$ defines a Toeplitz structure in the sense
of \cite{7,\rm p.18} (see the remark at the end of Ref. \cite{8}),
with underlying symplectic submanifold of
$\ T^*Q\setminus 0\ $ the positive cone over the graph of $\alpha$:
$$\Sigma=\{\;t\alpha(\lambda)\;|\;\lambda\in Q,\,t>0\ \}\ \subset\  T^*Q
\setminus 0\ .\tag 5-1
$$
(Here and in the following $\ T^*Q\setminus 0\ $ denotes the
total space $\ T^*Q\ $ with the zero section removed.)
Let $\tau_\Sigma:\Sigma\to M$ denote the natural projection.
A (global) Toeplitz operator of order $k$ associated to
$(\Sigma,\Pi)$ is by definition an operator $A:\Cal H\to\Cal H$ of the form
$\ A=\Pi\, R\,\Pi\ $, where $ R$ is a pseudo-differential operator of
order $k$. The principal  symbol of $A$ is the
restriction of the principal  symbol of $R$ (which is a
function on $T^*Q$) to $\Sigma$. It was shown in \cite{7} that
Toeplitz operators form a ring, and that the principal
symbol of Toeplitz operators is well defined and
obeys the same rules as for
pseudo-differential operators:
$$\gather
\sigma(A_1A_2)=\sigma(A_1)\sigma(A_2),\\
\sigma( [A_1,A_2])=\i\{\sigma(A_1),\sigma(A_2)\},
\endgather
$$
where the Poisson brackets are computed with respect to the
symplectic structure on $\Sigma$.

The generator of the circle action
$\frac {1}{\i}\frac {\partial}{\partial\varphi}$ gives a first order
Toeplitz operator
$D_\varphi$ with symbol
$\sigma(D_\varphi)(t\a(\l))=t$.
$D_\varphi$ operates on $\Hm$ as multiplication by $m$.
For $f\in \Cal P(M)$ let $M_f$ be the multiplication operator on
$\Lq$ and $T_f=\Pi\, M_f\,\Pi$. The symbol of $T_f$ is the pullback
of $f$ to $\Sigma$.
Being invariant under the
circle action, $T_f$ splits into a direct sum
$\ T_f=\oplus_{m=0}^\infty\Tfm\ .$
Identifying $\Hm$ with the space of holomorphic sections, the
operator $\Tfm$ on $\Hm$ is just the Toeplitz quantization (multiplication)
corresponding to
$f$ considered in the previous section.

\demo{Proof of Theorem 4.2}
The commutator
$[T_f,T_g]$ is a  Toeplitz operator of order $-1$
with principal symbol
$\ \i\{\tau_\Sigma^* f,\tau_\Sigma^*g\}_\Sigma(t\alpha(\lambda))=
\i\,t^{-1}\{f,g\}_M(\tau(\lambda))\ $.
It follows that the $S^1$-invariant, first order Toeplitz
operator
$$ A:=\,D_\varphi^2\,[T_f,T_g]-\i D_\varphi\, T_{\{f,g\}}
$$
has vanishing principal symbol, i.e. is in fact zeroth order.
But zeroth order pseudo-differential operators on compact manifolds
are bounded (see e.g\. \cite{16,\rm p.29}, or \cite{22}), and
since $\Pi$ is bounded as an operator on $\Lqv$,
it follows that $A$ is bounded. Since $\ ||A^{(m)}||\le||A||\ $ and
$$A^{(m)}=A|{\Cal H}^{(m)}=m^2[T_f^{(m)},T_g^{(m)}]-
\i\,m\,T_{\{f,g\}}^{(m)},
$$
we are done.\qed
\enddemo
\remark{Remark}
In a similar fashion, the theory in \cite{7} leads to
\roster
\item
Let $f\in \Cal P(M)$,\  $U^{(m)}(t)=\exp(-\i\, m\, t\,\Tfm)\ $ the
corresponding time evolution operator, and
$g\in C^\infty(M)$. If $F^t$ denotes the Hamiltonian flow for
$f$, one has
$$||U^{(m)}(t)\Tgm U^{(m)}(-t)-T^{(m)}_{(F^t)_*g}||=O(m^{-1})
\quad(\text{for }m\to\infty)\ .$$
This follows from the Egorov theorem for Toeplitz operators, see
\cite{7,\rm p.100}.
\item
For all $f_1,f_2,\ldots,f_r\in C^\infty(M)$,
$$||T^{(m)}_{f_1\ldots f_r}-
T^{(m)}_{f_1}\cdots
T^{(m)}_{f_r}||=O(m^{-1})\quad(\text{for }m\to\infty)\ .$$
\item
For all $f_1,f_2,\ldots,f_r\in C^{\infty}(M)$,
$$
\frac 1{\dim{\Cal H}^{(m)}} \text{tr}\,(T^{(m)}_{f_1}\cdots T^{(m)}_{f_r})
        =\frac 1{\text{vol}(M)}\int f_1\cdots f_r\; \Omega +O(m^{-1})\ .
$$
For the proof, see Guillemin \cite{18}.
\endroster
\endremark
\vskip 0.3cm
\demo{Proof of Theorem 4.1}
The second inequality is obvious. To prove the first,
we have to construct a sequence $\phm\in{\Cal H}^{(m)}$
such that
$$ \frac{||T_f^{(m)}\phm||}{||\phm||}=||f||_\infty+O(m^{-1})\ .
\tag 5-2 $$

The idea is to regard the $\phm$ as Fourier modes
(with respect to the $S^1$ action) of a single
distribution $\varPhi
\in{\Cal D}'(Q)$.
Let $x_0\in M$ be a point where
$|f(x_0)|=||f||_\infty $, and let $\lambda_0\in \tau^{-1}(x_0)$
be fixed. For $\lambda\in Q$, let
$$ \Xi_\lambda\ :=\ \{\;t\alpha(\lambda)\in T^*Q\;|\ t>0\ \}\tag 5-3$$
be the ray through $\alpha(\lambda)$.

We will look for a suitable $\varPhi$ among those distributions which have a
singularity at $\lambda_0$ in the direction of $\ \alpha(\lambda_0)\ $,
i.e\. whose
wave front set \cite{21} is contained in $\Xi_\lambda$ for
$\lambda=\lambda_0$.
A class of distributions having this property is the space $\ I^r(Q,\Xi)\ $
of ``Hermite distributions'' studied in \cite{7},\cite{19}:
Choose local coordinates $\ y=(y_1,\ldots,y_q),\,\,q=\dim Q\ $ around
$\lambda$ such that,
in the corresponding cotangent coordinates $\ (y,\eta)\ $,
the ray $\Xi_\lambda$ is given by the equations
$\ y_1=\ldots= y_q=0, \eta_2=\ldots \eta_q=0,\eta_1>0\ $.
Let us write $\ y'=(y_2,\ldots,y_q),\,\,\eta'=(\eta_2,\ldots,\eta_q)\ $.
Then the space $\ I^r(Q,\Xi_\lambda)\ $
consists of distributions $\varPhi$ that can be written, mod $C^\infty(Q)$,
as oscillatory integrals
$$
\varPhi(y)=(2\pi)^{-q}\int e^{\i y\eta}
a(\eta_1,\frac{\eta'}{\sqrt{|\eta_1|}}) d^q\eta\ .
\tag 5-4
$$
Here the amplitude $a(\eta_1,\eta')$ is smooth, vanishes for
 $\eta_1<\epsilon$ for some $\epsilon >0$,
and admits an asymptotic expansion
$$
a(\eta_1,\eta')\sim\sum_{j=0}^\infty a_j(\eta_1,\eta')
\tag 5-5$$
where $a_j$ is positively homogeneous
of degree $r-\frac{j+q}{2}$ in
$\eta_1$ for $\eta_1\gg 0$ and
 a Schwartz function in $\eta'$. It can be shown that this
definition does not depend on the particular choice of coordinates.
In particular, we can assume that
$\frac{\partial}{\partial y_1}=\frac{\partial}{\partial \varphi}$.

{}From  \cite{7}, Theorem 11.1 and 9.4, $\ I^r(Q,\Xi_\lambda)\ $
is invariant under
the Szeg\"o projector $\Pi$
and under zeroth order pseudo-differential
operators. In particular, it is invariant under $M_f$,
hence also under the Toeplitz operator $T_f$. Using that $f$ has a critical
point at $x_0$, the transport equation (\cite{7}, Theorem 10.2) shows
that
$$ (f-f(x_0))\varPhi\in I^{r-1}(Q,\Xi)
 \quad\text{for}\ \varPhi\in I^{r}(Q,\Xi)
 \ .\tag 5-6$$
We will need the following Lemma:
\proclaim{Lemma 1}
For all $\varPhi\in I^r(Q,\Xi_\lambda)$,
the Fourier modes $\phm$ have finite norm
admitting an asymptotic expansion
$$ ||\phm||^2\sim\sum_{j=0}^\infty b_j\,\, m^{2r-\frac{q+j+1}{2}}\tag 5-7$$
for $m\to\infty$
and vanish faster than any power for $m\to -\infty$.
Moreover, the leading term $b_0$ depends only on the equivalence class
in $I^r(Q,\Xi_\lambda)/I^{r-\frac{1}{2}}(Q,\Xi_\lambda)$, i.e. on its
``principal symbol''.
\endproclaim

Let us postpone the proof of Lemma 1 for a moment, and explain how
to make a particularly nice choice for $\Phi^{(m)}$.

Let $T_\lambda^*Q$ be the cotangent fiber. Since $T_\lambda^*Q\cap\Sigma
=\Xi_\lambda$, Theorem 9.4 from \cite{7} shows that $\Pi$ maps the space
$I^r(Q,T_\lambda^*Q-\{0\})$ of Fourier integrals into the space
$I^r(Q,\Xi_\lambda)$. Applying this to the delta function
$\delta_\lambda\in I^{q/2}(Q,T_\lambda^*Q-\{0\})$, we get some
$e_\lambda=\Pi\delta_\lambda\in I^{q/2}(Q,\Xi_\lambda)$. The Fourier
modes $e_\lambda^{(m)}$ of $e_\lambda$ have finite norm according to
Lemma 1, so they are in $\Hm$, and they satisfy for all
$\Psi^{(m)}\in\Hm$
$$ \langle e_\lambda^{(m)}|\Psi^{(m)}\rangle
    =\langle \delta_\lambda|\Pi^{(m)}|\Psi^{(m)}\rangle
    =\Psi^{(m)}(\lambda),\tag 5-8 $$
where again we have identified sections of $L^m$ with equivariant
functions.
On the other hand, (5-8) characterizes the $e^{(m)}_\lambda$
by Riesz' Lemma, and in fact (5-8) is used as
by Rawnsley \cite{28} as the defining property of his
``coherent states''.

\proclaim{Lemma 2}
For all $\lambda\in Q$,
$$ ||e_\lambda^{(m)}||^2=(2\pi)^{-n}m^n+O(m^{n-1/2}).$$
\endproclaim

\demo{Proof}
According to Lemma 1, the leading term depends only on the
principal symbol of $e_\lambda$. As for any statement concerning
principal symbol, it is therefore admissable to check the claim
in a ``model situation''. Model $Q$ as the unit circle bundle
in the tautological line bundle over $\P^n$. In this model, the
coherent states are explicitly known, and their squared norm
is $||e^{(m)}||^2=(2\pi)^{-n}(m+n)!/n!$ (see e.g. \cite{28}),
in accordance with the statement of the lemma.
\qed\enddemo

Let us now choose $\Phi=e_{\lambda_0}$. The two
lemmas (together with (5-6) show that
$$ \frac{||T_f^{(m)}\phm-f(x_0)\phm||}{||\phm||}=O(m^{-1}).$$
But this clearly gives (5-2) by the triangle inequality.
\qed

\remark{Remark}
The fact that the coherent states $e^{(m)}_\lambda$ are Fourier
modes of a Hermite distribution, together with Lemma 1, may
be used to derive a number of their asymptotic properties by
microanalytic means. For example:
\roster
\item
If $\tau(\lambda)\not=\tau(\mu)$, then
$$\langle e^{(m)}_\lambda,e^{(m)}_\mu\rangle=O(m^{-\infty}),$$
i.e. the coherent states are ``peaked'' at their base point.
\item
Let $f\in \Cal P(M)$ and  $U^{(m)}(t)=\exp(-\i\, m\, t\,\Tfm)\ $ the
corresponding time evolution operator.
If $F^t$ denotes the Hamiltonian flow for
$f$, one has
$$   \frac{   ||U^{(m)}(t)e^{(m)}_\lambda-e^{(m)}_{F^t(\lambda)}||}
          {||e^{(m)}_\lambda||}=O(m^{-\frac{1}{2}}),$$
i.e. the coherent states move according to the laws of classical
mechanics.
\endroster
\endremark

\vskip 0.3cm
\demo{Proof of Lemma 1}
Consider the following distribution on $S^1$:
$$
w(\varphi)=\sum_{m=-\infty}^\infty e^{\i m\varphi} ||\phm||^2=
\langle \varPhi|e^{\i\varphi D_\varphi}|\varPhi\rangle.
\tag 5-9$$
Since the singular support of $\ \varPhi\ $ is $\lambda_0$
 and the singular support
of $\ e^{\i\varphi D_\varphi}\varPhi$ is $\ e^{\i\varphi}\lambda_0\ $,
the distribution  $w$ is well-defined and
smooth away from $\varphi\in 2\pi{\Z}$. Let us study the singularity at
$\varphi=0$. (We may disregard the smooth part because the Fourier components
of a smooth function on $S^1$ go to zero faster than any power.)
Using the above local coordinates, one computes (mod smooth terms), using
Parseval's identity
$$\align
  w(\varphi)&=\int_Q \overline{\varPhi(\lambda)}\varPhi(e^{\i\varphi}
  \lambda) d\nu(\lambda)
         =(2\pi)^{-q}\int  e^{\i\varphi\eta_1}
         \big|a(\eta_1, \frac{\eta'}{\sqrt{|\eta_1|}})\big|^2 d\eta\\
         &=(2\pi)^{-q}\int  e^{\i\varphi\eta_1} |\eta_1|^{\frac{q-1}{2}}
          \bigg(\int  |a(\eta_1,\eta')|^2 d\eta'\bigg) d\eta_1\ .
\endalign $$
Since
$$ g(\eta_1)=(2\pi)^{-(q-1)}|\eta_1|^{\frac{q-1}{2}}
\int  |a(\eta_1,\eta')|^2 d \eta'
$$
is a classical symbol of order
$\frac{q-1}{2}+2(r-\frac{1}{4})=2r+\frac{q}{2}-1$
in the sense of H\"ormander, this is a classical
Fourier integral of order $2r+\frac{q}{2}-\frac{3}{4}$.
The full distribution is mod $C^\infty$
$$
w(\varphi)=(2\pi)^{-1}\sum_{k\in \Z}\int
e^{\i \tau(\varphi+2\pi k)}g(\tau) d\tau.
$$
With Poisson's summation formula,
this can be rewritten as a sum over the Fourier transforms:
$$w(\varphi)=\sum_{m\in \Z}g(m)e^{\i m\varphi}.$$
This shows that $\ ||\phm||^2=g(m)$ mod $m^{-\infty}$.
The Lemma now follows using the asymptotic expansion of the symbol $g$.
\qed
\enddemo
\enddemo

\head
Acknowledgement\hfill{ }
\endhead
We are very much indebted to J.B.~Bost for useful hints and
discussion. In particular, he suggested the use of the
global Toeplitz operators.


\enddocument
\bye